\documentclass[11pt]{article}

\usepackage{epsfig,multicol,bbm}
\usepackage{cite}
\usepackage{amsmath,amsfonts,amssymb}
\input epsf.sty

\usepackage{graphicx}
\usepackage{epstopdf}
\usepackage[body={17.5cm, 24cm},right=2.2cm]{geometry}
\usepackage{amssymb}
\usepackage{amsmath}

\newcommand{\bright}{\begin{flushright}}
\newcommand{\eright}{\end{flushright}}
\newcommand{\bminip}{\begin{minipage}}
\newcommand{\eminip}{\end{minipage}}
\newcommand{\bcent}{\begin{center}}
\newcommand{\ecent}{\end{center}}

\def\be{\begin{equation}}
\def\ee{\end{equation}}
\def\bea{\begin{eqnarray}}
\def\eea{\end{eqnarray}}

\def\nn{\nonumber}

\begin{document}

\thispagestyle{empty}

\begin{flushright}
IPPP/08/60\\
DCPT/08/120\\
UB-ECM-PF/08/18\\
\end{flushright}
\vskip 2cm

\begin{center}
{\Large \bf Warped Wilson Line DBI Inflation }
\end{center}
\vspace*{5mm} \noindent
\vskip 0.5cm

\date\today


\centerline{ \large A.~Avgoustidis$^a$~\footnote{
        E-mail address: tasos@ecm.ub.es} and
       I.~Zavala$^{b,c}$~\footnote{
        E-mail address: zavala@th.physik.uni-bonn.de}}

\vskip 1cm

\centerline{$^{a}$ \it Departament ECM and ICCUB, Universitat de Barcelona,}
\centerline{\it Diagonal 647, 08028, Barcelona, Spain  }
\vskip0.5cm
\centerline{$^{b}$ \it Institute for Particle Physics Phenomenology (IPPP)}
  \centerline{\it South Road, Durham DH1 3LE, United Kingdom}
  \vskip0.5cm
\centerline{$^{c}$ \it Bethe Center for Theoretical Physics and}
\centerline{\it Physikalisches Institut der Universit\"at Bonn}
\centerline{\it Nussallee 12, 53115 Bonn, Germany}
  \vskip2cm

\begin{abstract}
We propose a novel inflationary scenario in string theory in which the inflaton field is a {\em Wilson line} degree of freedom in the worldvolume of a probe D$p$-brane, in a warped flux compactification.  Kinetic terms for Wilson line fields on the world volume of a D-brane take a nonstandard Dirac-Born-Infeld (DBI) form. Thus, we work in the framework of DBI inflation. This extends the original slow roll Wilson line  inflationary scenario, where only the quadratic piece was considered.
Warped DBI Wilson line inflation offers an attractive alternative to ordinary (position field) DBI  inflation, inasmuch as observational and theoretical constraints get considerably relaxed.  Besides the standard large non-Gaussianities in DBI scenarios, it is also possible to achieve an observable amount of gravitational waves.

 \end{abstract}

\thispagestyle{empty}



\newpage

\tableofcontents

\section{Introduction}

Recent years have seen remarkable developments in the program of constructing cosmological models of inflation within string theory (for recent reviews see \cite{review,ms}).  However, we are still far from having explored all possible constructions and further possibilities continue to appear.
Warped D-brane inflation, where a brief period of accelerated expansion of the universe is  produced by the motion of a D-brane propagating in a warped throat, has received considerable attention \cite{review,ms, wsr,st}.  In particular the Dirac-Born-Infeld (DBI) inflationary scenario \cite{st} constitutes an attractive model with characteristic  features. In this model, a probe D-brane propagates relativistically in a warped compactification \'a la GKP \cite{KS,GKP}  (that is, internal fluxes are switched on, such that most of the closed string moduli get stabilised \cite{KKLT} and a warped throat is formed).
The brane dynamics is described by the DBI action, which contains non standard kinetic terms for the scalar fields associated to the brane's positions, acting as the inflaton(s)  \cite{st,dbi,bmca,LH,egmtz,shiu,wrappedbranes,hltw,AL,langlois}. A particular feature of the DBI scenario is that the sound speed of the fluctuations in the inflaton, can be much less that the speed of light, producing potentially large non-Gaussian signal in the density perturbations \cite{st}.

However, position DBI inflation has been put under severe theoretical and observational consistency constraints recently \cite{ms,bmca,LH,egmtz,AL,chen}. It was pointed out in \cite{bmca,LH,AL}, that inconsistent constraints on the scalar to tensor perturbations are obtained for single position field DBI models, both for D3-branes or wrapped branes. On the other hand, theoretical  backreaction constraints associated to the brane's high speed motion and the deformation of the background geometry, encoded in the warp factor, have been pointed out in \cite{ms,st,bmca,egmtz,chen}.

In this paper we discuss a different DBI scenario, where the consistency bounds can be considerably relaxed and moreover, potentially observable gravitational waves can be generated.
We start from the observation that in string compactifications, there are in general other open string moduli, besides position fields. In particular,  the so called Wilson Line (WL) moduli can be present. These fields have been widely used in phenomenological attempts  to construct realistic models of particle physics in  string theories \cite{wlpheno}.  In type II string theory, Wilson line degrees of freedom  can be turned on the world volume of D$p$-branes, along their internal wrapped $(p-3)$ dimensions.
Wilson line fields appear in the DBI action describing D$p$-branes as constant gauge fields, which have  interesting physics when turned on non contractible loops.  Thus, just as position moduli, they acquire nonstandard DBI kinetic terms.  It is then natural to ask, how the DBI picture changes, when the inflaton fields are identified not with position fields, but with WL modes, which  can also be present in  warped compactifications.
In such  a picture, the source of supersymmetry breaking is provided by a constant magnetic flux living on the world volume of the D-brane.
It is known that Wilson lines are T-dual to brane positions, whereas magnetic fluxes are dual to branes at angles \cite{polchis,clifford}.
Therefore, an interesting consequence of having WL's playing the role of inflatons in a warped compactification is that in principle, such a model can be T-dualised to a model with D-branes at angles, where brane  positions play the role of the inflaton. Thus, it provides a warped version with (partial) moduli stabilisation, of the models studied earlier on in \cite{branes@angles}.
Moreover, WL have already been used as inflaton fields in a slow roll Wilson Line inflationary model in \cite{acq}. In that case, one concentrates only in small derivatives, expanding the DBI action for the WL, which produces standard kinetic terms for the WL fields.

In this paper we study cosmological and theoretical constraints of DBI Wilson line inflation in a warped background, where most closed string moduli are stabilised by fluxes.  We call this scenario Warped Wilson Line DBI inflation.
We consider a probe D$p$-brane wrapping a $(p-3)$-cycle in a warped supergravity background. We then turn on Wilson lines as well as magnetic flux (as mentioned above, the magnetic flux has the job of breaking supersymmetry to generate a potential for the Wilson lines) on its world volume.
We concentrate in a single WL field, and assume that the open string moduli fields associated to the position of the D$p$-brane have been stabilised. At this stage we do not provide a precise mechanism to do this, but it could be realised along the lines of \cite{GomisMarchMat}\footnote{This technology has been applied to slow roll Wilson line inflation \cite{acq}.}.
We derive theoretical constraints as well as cosmological bounds on the tensor to scalar ratio for the WL DBI inflation. We find that, contrary to ordinary position DBI, Wilson line DBI inflation can be in agreement with theoretical approximations, as well as with observational constraints. Moreover,  there exists a region of theoretical parameters where warped WL inflation  can give a detectable gravitational wave spectrum, besides large non-Gaussian  perturbations\footnote{A class of position inflation models with these two properties was proposed recently in  \cite{hltw}.}.

We would like to stress that in our present investigation, we adopt a  phenomenological approach. Therefore, at this stage we  do not try to construct a concrete warped throat, which possess the required properties to embed a Wilson line DBI inflationary scenario. That is, such that  $(p-3)$-cycles exist, which contain (at least locally) non trivial one-cycles  where the Wilson lines can be turned on. Finding such geometries is  not a trivial task. However we believe that further developments of the theory will certainly provide the theoretical set up needed.

\smallskip

The structure of the paper is as follows.  Section~\ref{sec_setup} contains a general description of the warped Wilson line DBI set up.  We calculate the DBI and Wess-Zumino actions for general number of  position as well as Wilson line fields, and couple the system to gravity in order to study its cosmological implications.
In section~\ref{sec_cosmo} we explore the cosmological implications of single Wilson line field inflation.
We find theoretical constraints on the DBI Wilson line inflaton as well as cosmological bounds on tensor perturbations. We show that there exist regions of the  parameter space where all constraints can be satisfied, providing a novel model of DBI inflation which can produce observable non-Gaussianities as well as gravitational waves.
 We conclude in section~\ref{sec_concl}.

\section{General set up\label{sec_setup}}

In this section we derive the general action and equations
of motion for a wrapped probe D$p$ or anti-D$p$-brane moving
through a warped flux compactification in type IIB string theory.

We start by specifying the ansatz for the background fields and the form of the probe brane action.
In a warped flux  compactification in type IIB theory,
the ten dimensional metric takes the following general form (in the Einstein frame)
\be\label{metrica}
G_{MN}dx^Mdx^N= h^{-1/2}\,g_{\mu\nu}\,dx^\mu dx^\nu + h^{1/2}\,g_{AB} dy^A dy^B\,.
\ee

Here $h$ is the warp factor which depends on the compact coordinates\footnote{The warp factor is often a function of a single radial like coordinate but in more general configurations, it can also depend on the angular coordinates (see e.~g.~\cite{hek}).} $y^A$. The internal metric $ g_{AB}$ depends also on
the internal coordinates $y^A$.
In a general flux compactification in type IIB theory, all fluxes are turned on: RR forms $F_{n+1}=dC_n$ for $n=1, 3, 5$ and their duals $n=7, 9$ (remember also that $F_5$ is self dual), as well as NSNS flux $H_3=dB_2$.
These fields have only  compact components, thus the duals have legs in all four infinite dimensions plus the relevant components in the internal ones.
Besides these fields, the dilaton is in general active, and usually, it
is also a function of the six dimensional coordinates, $\phi=\phi(y)$.

We now embed a probe D$p$-brane (or anti-brane) in this background, which has four of its dimensions parallel to the four  infinite dimensions and $(p-3)$ spatial dimensions wrapped along an internal $(p-3)$-cycle. The motion of such a probe brane is described by the sum of the Dirac-Born-Infeld (DBI) and  Wess-Zumino (WZ) actions.  The DBI part is given, in the Einstein frame, by\footnote{
In the string frame, the DBI action is given by
$$ S_{DBI} = -\mu_p\,\int{d^{p+1}\xi\, e^{-\phi}
  \sqrt{- \det (\gamma_{ab}+ {\mathcal F}_{ab} )}}\,. $$
In D dimensions, the Einstein and string frames are related
by $G^E_{MN} = e^{-\frac{4}{(D-2)}\phi}G^s_{MN}$.  }
\be\label{dbi}
S_{DBI} = -\mu_p\,\int{d^{p+1}\xi\, e^{\frac{(p-3)}{4}\phi}
  \sqrt{- \det (\gamma_{ab}+ e^{-\frac{\phi}{2}}{\mathcal F}_{ab} )}}\,,
\ee
where the tension of a D$p$-brane in the Einstein frame is\footnote{Notice that
for a D$3$-brane, $T_p=\mu_p$ in the Einstein frame.}
$$T_p = \mu_p \,e^{\frac{(p-3)}{4}\phi} \qquad \quad
{\rm with} \quad \qquad \mu_p = (2\pi)^{-p}(\alpha')^{-(p+1)/2}\,. $$
Further, ${\mathcal F}_{ab}={\mathcal B}_{ ab}
+ 2\pi\alpha'\,F_{ab}$, with ${\mathcal B_2}$ the pullback of the NSNS 2-form field  on the brane,
$F_2$ the world volume gauge field and $\gamma_{ab} =  G_{MN}\,\partial_{a} x^M \partial_{b}  x^N$
the pullback of the ten-dimensional metric in the Einstein frame.  Finally, $\alpha'=\ell_s^2$ is
the string scale and $\xi^a$ are the brane world-volume coordinates.
Here, and in what follows, we label the coordinates using the following indices
\bea
&&  M,\,N = 0,1,\,\dots , 9 \qquad {\rm all \,\,\, 10D \,\,\, coordinates } \nonumber \\
&&   A,\,B = 4,\,\dots , 9 \qquad \quad\,\,{\rm all \,\,\, 6D \,\,\, coordinates } \nonumber\\
&& \mu,\,\nu = 0,1,\,\dots , 3 \qquad \,\,\,\,\,{\rm all \,\,\, 4D \,\,\, coordinates }\nonumber \\
&& a,\,b = 0,1,\,\dots , p \qquad \quad {\rm all \,\,\, worldvolume \,\,\, coordinates } \nonumber\\
&& m,\,n = 4,\,\dots , p \qquad \quad
 	\,\,\,{\rm internal \,\,\, (p-3) \,\,\, worldvolume \,\,\,coordinates
  					 }\nonumber   \\
&& i,\,j = p+1,\,\dots , 9 \qquad \,\,\, {\rm internal \,\,\, transverse \,\,\, to \,\,\,brane\,\,\,
     coordinates }\nonumber
\eea

The action (\ref{dbi}) is reliable for arbitrary values of the gradients
$\partial_a  x^M$, as long as these are themselves
slowly varying in space-time. That is, for small accelerations
compared to the string scale (equivalently, for small extrinsic
curvatures of the brane worldvolume). In addition, the string coupling
at the location of the brane should  be small in order for the perturbative expansion to hold, i.e.~$g_s\ll 1$, where $g_s = e^{\phi_0}$.

In a background with all fluxes turned on, a brane with nonzero two form flux on it, has a general  Wess-Zumino  action of the form%
\be\label{wz}
S_{WZ}= q\,\mu_p \,\int_{\mathcal W_{p+1}}\sum_n{\mathcal C}_n \wedge \,e^{\mathcal F}\,,
\ee
where ${\mathcal C}_n$ are the pullbacks of the background $C_n$ forms present, ${\mathcal F} = {\mathcal B} + 2\pi\alpha' F$ as defined before, and the  wedge product singles out the relevant terms in the exponential. Also ${\mathcal W_{p+1}}$ is the world volume of the brane and $q= 1$
for a probe D$p$-brane, while  $q=-1$ for a probe anti-brane.

In type IIB string theory with internal fluxes, the  contributions to the WZ action from D5 and D7 branes, to order  $(\alpha')^2$ are given by\footnote{Remember that the forms $C_8$ and $C_6$ are the duals of $C_0$ and $C_2$ which are present in a flux compactification.}
\bea\label{wz57}
&&\hskip-1cm
 S_{5WZ} = q\, \mu_5 \int_{\mathcal W_6} \left[ {\mathcal C}_6 +
		{\mathcal C}_4\wedge ({\mathcal B}_2 + 2\pi\alpha' F_2)
  \right], \nn \\
&& \hskip-1cm
S_{7WZ} = q\,\mu_7 \int_{\mathcal W_8} \left[ {\mathcal C}_8 +
	{\mathcal C}_6\wedge ({\mathcal B}_2 + 2\pi\alpha' F_2) +
   {\mathcal C}_4\wedge \left(\frac{1}{2}{\mathcal B}_2\wedge {\mathcal B}_2 +    {\mathcal B}_2\wedge 2\pi\alpha' F_2
     + \frac{(2\pi\alpha' )^2}{2} F_2\wedge F_2\right)\right]. \nonumber
\\
\eea

\subsection{Wilson Line DBI Kinetic Terms}

Before computing the determinant in (\ref{dbi}), we first review how Wilson line fields arise in the present set up.
Wilson lines correspond to constant background gauge potentials, $A_1 =$ const.~which can give interesting physics when turned on spaces of nontrivial topology.
Consider a constant gauge potential $A_1$, with corresponding field
strength $F_2$, subject to gauge transformations $A\rightarrow
A+{\rm d}\Lambda$ leaving $F$ invariant.  Given a closed path $\gamma$,  a Wilson line  is a gauge invariant quantity\footnote{Here we are only interested in the Abelian case.}, which  can be defined as a path ordered exponential of the  line integral of the gauge field along $\gamma$,  $U_\gamma={\rm P}\exp\oint_\gamma A \, , $
where $\rm P$ stands for path ordering. If $\gamma$ is contractible, one can use Stokes' theorem to trade the line integral along $\gamma$  for a surface integral over the region $C$ enclosed by $\gamma$:
$U_\gamma={\rm P}\exp\int_C F \,.$
Further, if  $C$ is simply connected, the Wilson line is trivial whenever the magnetic field is zero, $F=0$ ($U_\gamma=1)$.  However, when $C$ is not simply connected, one can have $F=0$ but $U_\gamma\ne 1$ in a gauge invariant fashion. Thus, in manifolds of nontrivial topology,  Wilson lines can give rise to physical effects, even if the actual  magnetic field is zero.

When turned on D$p$-branes wrapping $(p-3)$ internal dimensions in a compact six dimensional space, these degrees of freedom correspond to scalar fields from the four dimensional point of view. In order to see this, notice that the magnetic field living on the brane $F_{ab}$  has three types of components $F_{mn}$, $F_{\mu\nu}$ and  $F_{\mu n}$. We want to preserve Lorentz invariance in the four large dimensions, therefore we take $F_{\mu\nu}=0$. Further, a constant  $F_{mn}\ne0$ corresponds to internal magnetic flux, which in general breaks supersymmetry. This can be easily understood from a T-dual picture where  D-branes with  magnetic flux  correspond to branes at angles \cite{polchis,clifford}. For a generic set  of angles, supersymmetry is broken \cite{douglas}.
 Nevertheless, $F_{mn} =0 $ can give rise to a Wilson line, if say $F_{mn} = \partial_m A_n-\partial_n A_m=0$, that is $A_{n}$  are independent of the internal coordinates and there exists a non trivial topology in  the $(p-3)$-cycle wrapped by the D$p$-brane (i.~e.~the first homotopy group of the cycle is nonzero).
Notice that  $A_{n}$ can still depend on the large four dimensions, thus giving rise to a non vanishing component $F_{\mu n}$.  This is, from the four dimensional point of view, the scalar field associated to the Wilson lines.
These fields do not break supersymmetry, which again can be understood from a T-dual point of view, as  they correspond to D$p$-brane positions \cite{polchis,clifford}. That is, they are open string  moduli of the theory.
In what follows, we consider  Wilson line fields\footnote{In a slight abuse of terminology we refer to $A_{n}$ as the Wilson line, instead of  the object $U_\gamma$. This should not cause any major confusion in what follows. } as well as constant internal magnetic fields on the probe brane, which break supersymmetry generating a potential for the WL's as  shown in \cite{acq}.

We work in static  gauge, identifying the non compact
worldvolume coordinates with the corresponding spacetime coordinates,
$\xi^\mu = x^{\mu}$, while allowing for the moment the $(p-3)$ background
coordinates wrapped by the brane to be functions of the compact
worldvolume coordinates only, $y^{m}=y^{m}(\xi^n)$.
We also allow  the background coordinates transverse to the brane
to be functions of the large four dimensional worldvolume coordinates, $y^i=y^{i}(\xi^\mu)$.
With these choices, the components of the induced metric on the brane worldvolume are:
\bea
 \gamma_{\mu\nu} = G_{\mu\nu} + \frac{\partial y^i}{\partial \xi^\mu}\,
	\frac{\partial y^j}{\partial \xi^\nu}\,G_{ij} \,, 
 \qquad \qquad
 \gamma_{mn} = \frac{\partial y^l}{\partial \xi^m}\frac{\partial y^r}{\partial \xi^n}\,G_{lr} \,.
 \label{gamma}
\eea

\noindent Similarly, the pullback of the NS two form field has components
\bea\label{pullB}
{\mathcal B}_{\mu\nu} = \frac{\partial y^i}{\partial \xi^\mu}\,
	\frac{\partial y^j}{\partial \xi^\nu}\, B_{ij}\,, 
\qquad \qquad
{\mathcal B}_{mn} = \frac{\partial y^l}{\partial \xi^m}\,
	\frac{\partial y^r}{\partial \xi^n}\, B_{lr} \,.
\eea

\smallskip

\noindent It is now clear that the matrix in the DBI action (\ref{dbi}) has the general form
\be
\gamma_{ab} + {\mathcal F}_{ab} = \left( \begin{array}{cc}
  \gamma_{\mu\nu} + e^{-\frac{\phi}{2}} {\mathcal B}_{\mu\nu}
         &  2\pi\alpha'e^{-\frac{\phi}{2}} F_{\mu n}  \\
 	 2\pi\alpha'e^{-\frac{\phi}{2}}F_{n\nu} & {E}_{mn}  \\
              \end{array}   \right) \,,
\ee
where we defined ${E}_{mn} = \gamma_{mn}+ e^{-\frac{\phi}{2}}({\mathcal B}_{mn}+2\pi\alpha' F_{mn})$. The four dimensional contribution to the pull back of the NSNS two form, ${\mathcal B}_{\mu\nu} $ (see eq.~(\ref{pullB})) will be  in general nonzero.
 For cosmological purposes, we will be interested in homogeneous configurations. In that case, it is easy to see that this term vanishes. Notice however that it will be important for the study of cosmological perturbations with multiple position fields when NSNS flux is present (both in slow roll and DBI scenarios). How this type of terms affect the resulting perturbation spectrum in the multifield case, will be discussed elsewhere \cite{tz}.
Since we will be mostly interested in a single field case in next section, from now on, we simply  ignore this term.

At this point, we can use the identity
\be\label{matrix_id2}
\det \left( \begin{array}{cc}
                  \bf A & \bf B   \\
                  \bf C & \bf D   \\
            \end{array}          \right)
= \det(\bf D) \, \det( {\bf A} - {\bf B} {\bf D}^{-1} {\bf C} ) \,,
\ee
to compute the determinant.  Using this expression and defining the inverse of ${E}_{mn}$ with upper indices such that ${E}_{mn}{E}^{nl} = \delta_m^{\,l}$, we find
\bea
 \det[\gamma_{ab} +{\mathcal F}_{ab}] &=& \det[{E}_{mn}] \, \det[G_{\mu\nu}
 + \partial_\mu y^i \partial_\nu y^j G_{ij} + (2\pi\alpha')^2e^{-\phi} F_{\mu n}{E}^{nm} F_{\nu m}] \, \nn \\
&=& h^{-2} \det[{E}_{mn}]  \, \det[g_{\mu\beta}] \, \det[
   \delta^{\beta}_\nu  + h \,\partial^{\beta}y^i \partial_\nu y^j g_{ij}
    + h^{1/2}(2\pi\alpha')^2e^{-\phi} \partial^{\beta} A_n\partial_{\nu}A_m {E}^{nm} ]\,, \nn\\
\eea
where we have used that $G_{\mu\nu} = h^{-1/2}g_{\mu\nu}$, $G_{ij} = h^{1/2}g_{ij}$ and $F_{\mu n}= \partial_\mu A_n$ (thus from now on contractions are done using $g_{MN}$).
Now we can use general expressions for the determinant of a $4\times 4 $ matrix to compute the last determinant in the previous equation\footnote{See e.~g.~Appendix 1 of the second reference in \cite{langlois}.} and express it as follows
\bea\label{FullDet}
 \det[
   \delta^{\beta}_\nu  + H_Y Y^\beta_I Y_\nu^I ]& =& 1
   + H_Y Y^\nu_I Y_\nu^I  + H_Y^2 Y^{[\nu}_I Y_\nu^I Y^{\beta]}_J Y_\beta^J + H_Y^3 Y^{[\nu}_I Y_\nu^I Y^{\beta}_J Y_\beta^J
   Y^{\lambda]}_K Y_\lambda^K    \nn \\
&&\hskip5cm   +  H^4_Y Y^{[\nu}_I Y_\nu^I
   Y^{\beta}_J Y_\beta^J Y^{\lambda}_K Y_\lambda^K
   Y^{\gamma]}_L Y_\gamma^L \, ,
\eea
where $ Y^\beta_I \equiv \partial^\beta Y_I $  with $Y^I = (y^i, \,A^m)$ and $H_{Y} = (h, \, h^{1/2}(2\pi\alpha')^2e^{-\phi})$. Also note that $H_Y^2$ in the above equation should be
understood as a shorthand for $H_{Y^I} H_{Y^J}$, and similarly for higher powers of $H_Y$.
Thus, this expression includes all fields present, and one must remember that contraction of $y$
is done with the metric $g_{ij}$, whereas that of $A$ is done with $E^{mn}$.  Antisymmetrisation
is done with respect to greek indices.  In order to see how this
expression looks in terms of the position and Wilson line fields, we write explicitly the
determinant for the two-field case (one position field $y$ and one Wilson Line A), keeping in mind
that when more fields are present, one should use equation~(\ref{FullDet}).

With these considerations in mind, we arrive at the general expression for the DBI action (remember that the internal pieces $(m,n)$ depend only on the compact dimensions)
\bea\label{dbifinal}
&& \hskip-0.7cmS_{DBI} = -\mu_p\,\int{d^{p-3}\xi \, e^{\frac{(p-3)}{4}\phi} \,\sqrt{\det(E_{mn})}}\,
	\int{d^4x \,h^{-1}\,\sqrt{-\det(g_{\mu\nu})}}  \nonumber \\
 && \hskip-0.5cm \times \sqrt{1+ h\,(\partial y)^2 +
  h^{1/2}(2\pi\alpha')^2 e^{-\phi}\,(\partial A)^2 +
  h^{3/2}(2\pi\alpha')^2 e^{-\phi}(\partial y)^2 (\partial A )^2
- h^{3/2} (2\pi\alpha')^2 e^{-\phi}\,(\partial y \cdot \partial A)^2 }\,.\nonumber \\
\eea
Taking further a  static gauge for all worldvolume coordinates, that is  $\xi^a = (x^\mu,\, y^n)$,  we simply have $\gamma_{mn} = h^{1/2} g_{mn}$ and also ${\mathcal B}_{mn} = B_{mn}$. Thus $E_{mn} = h^{1/2}g_{mn} + e^{-\frac{\phi}{2}}(B_{mn} + 2\pi\alpha'F_{mn})$.

\smallskip

At this point we can simplify the kinetic terms by observing that the contribution of the magnetic field to $E^{mn}$ vanishes, because the components $F_{mn}$ are always transverse to the Wilson lines. Further, we can consider a configuration such that
$E^{mn}$  in eq.~(\ref{dbifinal}) can be replaced by $E^{mn} \to h^{-1/2} g^{mn}$ (that is, such that the $B_{mn}$ contribution vanishes).  Taking this into account and defining the gauge potential as
\be
A_n = \frac{\lambda_n}{L_n}\,,
\ee
 where $L_n$ is the size of the 1-cycle where the
Wilson line is turned on, and $\lambda_n \in [-\pi,\, \pi]$, we end up with
\bea
&& \hskip-0.7cmS_{DBI} = -\mu_p\,\int{d^{p-3}\xi \, e^{\frac{(p-3)}{4}\phi} \,\sqrt{\det(E_{mn})}}\,
	\int{d^4x \,h^{-1}\,\sqrt{-\det(g_{\mu\nu})}}   \nonumber \\
 && \hskip-0.5cm \times \sqrt{1+ h\,(\partial y)^2 +
  (2\pi\alpha')^2 e^{-\phi}\left(\partial (\lambda/L) \right)^2 + (2\pi\alpha')^2 e^{-\phi}(\partial y)^2 \,(\partial(\lambda/L))^2
- (2\pi\alpha')^2 e^{-\phi}(\partial y \cdot\partial (\lambda/L))^2 }\,,\nonumber \\
\eea
in the two field case.
We can  now define  the canonically normalised fields associated to the brane positions and Wilson lines as follows:
\bea\label{canonics1}
&& f(\varphi) = h^{-1}\, \mu_p\,\int{d^{p-3}\xi \,e^{\frac{(p-3)}{4}\phi} \,\sqrt{\det(E_{mn})}}\,,\\
&&  d\varphi =  \left[ \mu_p\,\int{d^{p-3}\xi \, e^{\frac{(p-3)}{4}\phi} \,\sqrt{\det(E_{mn})}}
			\right]^{1/2}\,dy\label{canonics2} \,, \\
&& d\chi= \left[  \frac{(2\pi\alpha')^2 \,e^{-\phi}}{L^2}\,f(\varphi) \right]^{1/2} d\lambda  \,.
\label{canonics3}
\eea
Using these definitions, the DBI action for the position and WL fields  becomes,
\bea\label{canodbi}
S_{DBI}&=&\int{d^4x \,\sqrt{-g}} \,f(\varphi) 
    \nonumber \\
  &&\hskip1cm \times  \sqrt{1
	+ f(\varphi)^{-1}(\partial\varphi)^2
        + f(\varphi)^{-1}(\partial \chi)^2  + f(\varphi)^{-2}(\partial\varphi)^2(\partial \chi)^2  - f(\varphi)^{-2}(\partial \varphi\cdot\partial\chi)^2 } \,,
         \nonumber \\
\eea
where we have abbreviated $\det(g_{\mu\nu})\equiv g$ and remember that we are considering only two fields. However, this expression can be easily   generalised to the case when more than two fields are present.
What is important is to notice the mixing between position and WL fields. These terms  can give interesting effects when considering situations where both types of fields play a role in inflation.
Notice also the difference arising among the two types of fields, Wilson line moduli and position moduli, due to their different origin. The WL fields get a multiplicative factor that is the inverse of the internal background metric $g^{mn}$, which is of course, independent of the WL fields.  Instead, the position fields are contracted with the internal metric which can depend on all the position fields themselves.
This is also a difference with the flat Wilson line standard slow roll inflationary scenario \cite{acq}, in which the kinetic terms are multiplied by a flat metric.  We will see in Section \ref{sec_cosmo} how these differences change the cosmological implications of warped WL inflation.

\subsection{Wess-Zumino terms}

In this section we discuss the general form of the WZ action for a probe D$p$-brane moving in a warped flux compactification.  As the background geometry contains only internal fluxes such that four dimensional Lorentz invariance is preserved,  the Hodge duals of the fluxes have legs in the four large dimensions plus some internal components. More concretely, in type IIB theory the following internal fluxes $F_1$,  $F_3$, $F_5$, $H_3$ can be turned on. The RR fluxes have correspondingly Hodge  duals  $F_9$, $F_7$ and we know that $F_5$ is self dual\footnote{
Remember that the dual of an $n$-form field strength in $D$-dimensions is defined by
$$ ^\star F_n = \frac{\sqrt{-g_D}}{(D-n)!} \epsilon_{1\dots D} F^{D-q}
\,.$$}.
The general form for the dual background fluxes and the brane magnetic flux, can be expressed as follows:
\bea\label{fluxes}
&& B_2 = b(y) \,\omega_2 \nonumber  \\
&& F_2 = f_2 \, \tilde \omega_2\nonumber \\
&&   C_4 = N h^{-1}(y) \sqrt{-g} \,dx^0 \wedge \cdots \wedge dx^3 \nonumber\\
&&   C_6 = M \eta_6(y) \sqrt{-g} \,dx^0 \wedge \cdots \wedge dx^3   \wedge \alpha_2 \nonumber \\
&&   C_8 = K \eta_8(y) \sqrt{-g} \,dx^0 \wedge \cdots \wedge dx^3 \wedge \omega_4 \,,
\eea
where $\omega_2,\, \tilde\omega_2\,, \, \alpha_2$ and $\omega_4$ are respectively two  and four forms defined suitably in the internal dimensions and normalised such that
$$\int_{\Sigma_2} \omega_2 = \int_{\Sigma_2} \tilde\omega_2 =\int_{\Sigma_2}\alpha_2 =1,
\quad \qquad  \int_{\Sigma_4} \omega_4 = 1 \,,$$
where $\Sigma_i$ corresponds to an $i$-cycle in the internal dimensions.
 Also $f_2$ is a constant magnetic flux on the branes, which in general  breaks supersymmetry\footnote{Note however that,  when $H_3$ flux is present, a suitable magnetic field $F_2$ has to be turned on $D7$-branes to preserve supersymmetry in warped compactifications \cite{GomisMarchMat,cos}.},
and the constants  $N,\,M,\,K$ are the corresponding flux units.  Further, $b(y)$, $\eta_i(y)$ are some functions
of  the internal components and $h(y)$ is the warp factor (see eq.~(\ref{metrica})).
Plugging (\ref{fluxes}) into (\ref{wz57}), the WZ contributions boil down to  general expressions of the form:
\bea\label{finalWZ5}
&& S_{5WZ} = q\,\mu_5 \int{d^4x \sqrt{-g} \, \left[ M\eta_6(y)
	+N h^{-1}(y)\left( b(y) + 2\pi\alpha'  \,f_2 \right)  \right]}\,, \\
&& S_{7WZ} = q\,\mu_7  \int d^4x \sqrt{-g} \, \Big[ K \eta_8(y)
    + M\eta_6(y)\left(b(y) + 2\pi\alpha' \, f_2\right)  \nonumber \\
&&\hskip5cm	
 	+N h^{-1}(y)\left(\frac{1}{2}\,b(y)^2 + 2\pi\alpha'  \,f_2\, b(y)
	+ \frac{2\pi\alpha'}{2}  \,f_2^2\right)  \Big] \label{finalWZ7} \,.
\eea
Notice that in passing from geometric coordinates to  four dimensional scalar fields, $\varphi^i$, we have to re-scale the functions in (\ref{fluxes}) as  $\eta_5(\varphi^i) = \mu_5\eta_5(y)$, $h^{-1}(\varphi^i) = \mu_5 h^{-1}(y)$, and so on, so that the units are correct.  Notice also
 that the WZ terms are independent of the WL fields.

\subsection{Dynamics}

We now look at  the dynamics associated to the actions (\ref{canodbi}),  (\ref{finalWZ5}) and  (\ref{finalWZ7})
coupled to gravity.  Since we are interested in cosmological solutions,  from now on we consider  homogeneous fields $\varphi^i(t)$, $\chi_n(t)$, independent of spatial position to zeroth order\footnote{One must keep in mind  that the mixed terms in (\ref{FullDet}) (or (\ref{canodbi}) for two field case), will be important for the cosmological perturbations analysis when both types of scalar fields, positions and Wilson lines, are evolving. } and we take  a FLRW ansatz for the four dimensional metric $g_{\mu\nu}$.
Introducing further generic potentials
$V_\varphi(\varphi^i)$
and $V_\chi(\chi_n)$ for the canonically normalised fields associated to the positions of the branes and Wilson lines respectively, the
total 4D action reads:
\bea
S_4&\!=\!&\frac{M_{Pl}^2}{2}\int{d^4x \sqrt{-g} \, R }\, \nn\\
	&&\hskip-0.4cm -
        \int{d^4x \,\sqrt{-g}} \, \bigg[ f(\varphi) \sqrt{1
	- f(\varphi)^{-1}\dot \varphi^i\dot\varphi^j g_{ij}
        - f(\varphi)^{-1}\dot \chi_n \dot \chi_m \, g^{nm}}
        + V_\varphi(\varphi^i) + V_\chi(\chi_n)
        - q  F_p(\varphi) \bigg]
        \label{S_tot} , \nonumber \\
\eea
where the Planck mass is given in terms of the six dimensional volume as $M_{Pl}^2= V_6/\kappa_{10}^2$, with $\kappa_{10}^2= (2\pi)^7(\alpha')^4g_s^2/2$. Also, we have expressed the $p$-brane Wess-Zumino part in terms of a general function $F_p(\varphi)$.
The potentials introduced above will arise once coupling to other sectors and other effects, like moduli stabilisation, are taken into account. As we will see in the next section, the exact form of the potential is not relevant for our analysis. Therefore we do not take in any particular form for it\footnote{However, notice that
in the original Wilson line slow roll inflation model \cite{acq}, a potential of the form $V\sim -a^2/(b^2+\chi^2)^m$, with $a\propto f_2$ and $b\propto \varphi$, arising from the interaction between two parallel branes in the presence of (worldvolume) magnetic flux was found. Including the vacuum contribution, this gave  $V\sim \tilde a+\tilde b\chi^2$ for $\chi \ll b$ \cite{acq}.}.

  The equations of motion derived from the action (\ref{S_tot}) thus take the form
\bea
H^2 &=& \frac{1}{3M_{Pl}^2} \,\rho \,, \label{Fried}\\
 \dot \rho + 3H (\rho + p) &=&0 \,,\label{fluid}\\
 \frac{1}{a^3} \frac{d}{dt}\left[a^3\gamma\,g_{ij}\,\dot\varphi^j \right]  &=& -\frac{f_{,i}}{\gamma}
+ \frac{f_{,i}}{2\,f}\gamma\, v^2 + q \,F_{,i}+ \frac{\gamma}{2} \left[g_{kl,i}\,\dot\varphi^k\dot\varphi^l
    + g^{mn}_{\,\,\,\,\,\,\,,i}\dot\chi_m \dot\chi_n  \right] -
    \frac{\partial V_\varphi}{\partial\varphi^i} \,, \label{positioneq}\\
   \frac{1}{a^3} \frac{d}{dt}\left[a^3\gamma\,g^{mn}\dot\chi_n \right]  &=& - \frac{\partial V_\chi}{\partial \chi_m}  \,\,. \label{wleq}
\eea
Here we have defined a relativistic Lorentz factor as:
\be\label{gamma_Lor}
\gamma=\frac{1}{\sqrt{1-f^{-1}v^2}}\,,
\ee
characteristic of DBI dynamics, where
$v^2=\dot\varphi^i\dot\varphi^j g_{ij}+ \dot\chi_m\dot\chi_n g^{mn}$.
$H\equiv \dot a/a$ is the Hubble variable, $a=a(t)$ being the scale factor of the FLRW background. The energy
density and pressure are obtained from the nonzero components of the  energy-momentum tensor, which
take a perfect fluid form with
\bea
T_{00} =\rho &=& f\gamma - q\, F_p + V\label{T00}\,, \\
T_{ii} = p &=& q\, F_p - f\gamma^{-1} - V
\hskip0.5cm ({\rm no\ sum\ in\ } i), \label{Tii}
\eea
where we have defined the total potential $V(\varphi,\chi)=
V_\varphi(\varphi^i)+V_\chi(\chi_n)$.
These give rise to the (variable) equation of state:
\be\label{eos}
w\equiv \frac{p}{\rho} = \frac{-\gamma^{-2}-(Vf^{-1}-q\,f^{-1}F_p)
\gamma^{-1}}{1+(Vf^{-1}-q\, f^{-1}F_p)\gamma^{-1}} \, ,
\ee
and  speed of sound
\be
c^2_s = \frac{\partial p}{\partial \rho} = \gamma^{-2} \,.
\ee

Before concluding this section some comments  are in order.  In the discussion above, we have implicitly assumed that the constant magnetic field and the Wilson lines, are turned on the same brane. However it is possible to separate the {\em supersymmetry breaking brane} form the {\em inflationary brane}, as was done in the slow roll WL model
\cite{acq}. In that case, one brane is responsible for generating the potential for the WL inflaton field  through a constant magnetic field turned on its world volume, whereas a second brane  drives inflation via Wilson lines.
The discussion above carries over in such a case, by simply taking $f_2=0$ on the inflationary brane.
In the rest of the paper, we keep the discussion general by allowing the magnetic and Wilson lines to live on the same brane, but one should keep in mind that the two-brane picture just discussed, is also included.

 In what follows, we study the cosmological implications of the system discussed in the present section.

\section{Wilson Line DBI Inflation\label{sec_cosmo}}

In this section, we investigate the consequences of having a Wilson line field playing a role in
cosmology in a warped set up and, in particular, its implications for inflation.  In order to extract the phenomenology of warped Wilson line inflation, we concentrate
on the case in which only these fields are responsible for the cosmological evolution
of the four dimensional spacetime.  Thus we assume  that  the positions of the wrapped branes have
been fixed by some mechanism\footnote{Open string moduli like brane positions can be stabilised in the presence of $G_3$ flux and brane magnetic flux, by using the technology developed in reference
\cite{GomisMarchMat}. This has been applied to slow roll Wilson line inflation in \cite{acq}.}.

Under these considerations, the action in four dimensions relevant for cosmology is (in this section we consider only D-branes, that is $q=1$)
\bea
S_4= \frac{M_{Pl}^2}{2}\int{d^4x \sqrt{-g} \,R } - \int{d^4x \sqrt{-g}\,\left[f_0\,\sqrt{1-
	f_0^{-1}\,\dot\chi_n\,\dot\chi_m \, g^{mn}}
 			+ V(\chi_n) + V_{\varphi_0} + F_{0_p}\right]} \,.
\eea
Here we have denoted $f_0 = f(\varphi_0)$ with $f(\varphi)$ given in (\ref{canonics1}) and
$\chi_n$ the Wilson line field, defined in (\ref{canonics3}) with $e^{-\phi_0}= g_s^{-1}$.
$V_{\varphi_0}$ is the potential for the position field(s), evaluated at the stable  point $\varphi_0$. Further, $F_{0_p}$  comes from the WZ term  (\ref{finalWZ5}-\ref{finalWZ7}), which when evaluated at a fixed position, becomes simply a constant. This constant depends on the fluxes and warp factor and can be seen as a constant piece added to the Wilson line potential, together with $V_{\varphi_0}$.

Just as in standard DBI inflation, there is a `speed limit', $\dot\chi_n\,\dot\chi_m g^{mn} < f_0 $, coming from the nonstandard form of the Wilson line kinetic terms.  However, in the present case, the time derivatives of the WL fields are not associated directly to brane positions (although indirectly  they are, in a T-dual picture). In other words, there is no motion of the wrapped probe brane inside the throat.
Still, there are some backreaction issues which one needs to take into account.

\subsubsection*{Backreaction issues}

There are two  backreaction issues one has to take care of in order to determine the region of parameters where our  approximations are consistent. First of all, for a probe
D-brane approximation to be valid, we have to ensure that the  background geometry  does not get distorted when we place a probe brane on it. This can be easily achieved in a warped flux compactification by taking  large values of flux, where the supergravity approximation is valid (the curvatures go like $\sim (flux)^{-1}$). Thus, placing a single D$p$-brane  will not  affect the geometry in a considerable way.

The second issue concerns the backreaction due to the WL approaching the ``speed limit"\footnote{We thank R.~Gregory for discussions on this point. }. We can estimate this effect  in terms of a T-dual picture where the WL is mapped to a brane location. Therefore, this gives rise to the same type of bound associated to standard DBI inflation, as discussed in detail in \cite{egmtz}.
We can see this as follows. Consider a D$5$-brane (the same reasoning can be used for a D$7$-brane) spanning  the four large dimensions and wrapping along the $y^4$ and $y^5$ internal dimensions,  in an AdS$_5\times$S$^5$ background, where only $F_5$ flux is present. Now turn on a Wilson line along the $y^4$ direction say (assuming that this is possible in some way). We can now T-dualise twice along $y^4$ and $y^5$ to get back to a picture of a  D$3$-brane moving at high speed. In this dual picture, we can follow the toy model backreaction calculation of \cite{egmtz}. Thus we obtain that, in order for the backreaction to be under control, we need \cite{egmtz}
\be\label{backreaction1}
 \gamma(\chi) \ll \frac{R^4}{g_s\ell_s^4} \,,
 \ee
where $R^4$ gives (the inverse of) the curvature of the ambient space, and results  proportional to the fluxes in a warped background. For example, in a Klebanov-Strassler (KS) background, it is proportional to $(g_sM)^2$ \cite{KS}. Therefore, once again, we require large values of flux for our approximations to be valid.
This condition was hard to achieve in a successful scenario in the standard position DBI  cases (see \cite{st,LH,egmtz,wrappedbranes,AL} for more details).

The attentive reader will have noticed a flaw in the estimation above. Indeed, in a more realistic flux compactification as we are interested in, all fluxes are turned on. In particular, an NSNS $H_3$ flux. It is known that, when such flux is present, T-duality mixes the metric with the $B_2$ field \cite{buscher}, producing a more complicated geometry. For example, a toroidal compactification with $B_2$ flux gives rise to twisted tori in the T-dual version \cite{micu,ttori}. For the more complicated case of a Calabi-Yau (CY), the T-dual geometry is not a CY manifold anymore \cite{micu}.
However, AdS$_5\times$S$^5$ serves as a good toy model to estimate the order of magnitude of the backreaction effects in our scenario, which is our main concern here. We expect the same type of relation between the Lorentz contraction factor and the ambient curvature to hold also in  more complicated geometries.

\subsection{Wilson Line Field Range}

In order to extract all relevant information on how  Wilson line DBI inflation works we concentrate on single WL field case in what follows. In this situation the Lorentz parameter reduces to
\be
\gamma = \frac{1}{\sqrt{1-f_0^{-1} g^{\chi\chi}\dot\chi^2}}\,.
\ee
Here $g^{\chi\chi}$ refers to a generic metric component associated to the direction along which the WL is turned on. Notice that this depends on the background coordinates only, i.e.~the brane positions. As we are assuming that these have been stabilised, $g^{\chi\chi}$ is just a constant. Its precise value is model dependent and will change according to the background geometry under consideration.  Notice further that this constant has to be included in the canonical normalisation of the WL field. Therefore from now on, we redefine  the WL field as $\sqrt{g^{\chi\chi}} \chi \to \chi$.

In order to determine the range of the WL field in Planck units, we have to  look at its associated canonically normalised field. From our definition in (\ref{canonics3}), we can see that the canonically normalised WL field is given
by (remember that we have included the geometric factor inside $\chi$):
\be
\chi = \left[ g^{\chi\chi}g_s^{-1}f_0 \frac{(2\pi\,\alpha')^2}{L^2}\right]^{1/2} \,\lambda \,.
\ee
Remember that $\lambda \in [-\pi,\pi]$, $\alpha' = \ell_s^2 = M_s^{-2}$ is the string scale, $L$ is  the size of the 1-cycle where the WL is turned on, and $f_0$ is given by (we discuss the explicit form of  this parameter below)
\be
f_0 = h_0^{-1} \,g_s^{(p-3)/4}\mu_p \int{d^{p-3}\xi\,\sqrt{det(E_{0\,mn})}} \, ,
\ee
with $\mu_p= (2\pi)^{-p}M_s^{p+1}$ and all quantities are evaluated at fixed values of the position fields $\varphi^i_0$. Furthermore, $f_0$ has units of $[mass]^4$.

We want to extract general statements regarding the field range  in terms of generic properties of the warped compactifications, in agreement to our approximations. For this, it is convenient to work with dimensionless quantities in string units. For example, we define the dimensionless six dimensional volume, in string units as
${\mathcal V_6} \equiv V_6/\ell_s^6$.

Let us start by rewriting the Planck mass in terms of purely dimensionless quantities and the string scale as follows:
\be\label{planckmass}
M_{Pl}^2  = \frac{2}{(2\pi)^7}\frac{\mathcal V_6}{g_s^2}\,M_s^2  \,.
\ee
Now we rewrite the WL field in terms of dimensionless parameters (in string units) as
\be
\frac{\chi^2}{M_{Pl}^2} = \frac{(2\pi)^9}{2}\,g^{\chi\chi} g_s\
                 \frac{{\mathfrak f}_0}{l^2\,{\mathcal V_6}} \, \lambda^2 \,.
\ee
Here we have defined ${\mathfrak f}_0 \equiv f_0/M_s^4$ and  $l\equiv L/\ell_s$.
This expression provides the range of the WL field (in Planck units) in a warped compactification in terms of dimensionless parameters (in string units).  String perturbation expansion requires that we take $g_s<1$. Moreover, in order for stringy effects to be negligible (such that the 4D approximation is valid), compactification scales have to be large in string units. That is, we require that $l>1$ as well as ${\mathcal V_6}^{1/6}>1 $. The value of $g^{\chi\chi}$ is model dependent (and dimensionless in our conventions).
Finally since the maximum value of $\lambda = \pi$, we have that during inflation the range of $\chi$ is bounded by
\be\label{dimlesschi}
\left(\frac{\Delta \chi}{M_{Pl}}\right)^{2} < 2^8\pi^{11}\,g^{\chi\chi} g_s\,
                 \frac{{\mathfrak f}_0}{l^2\,{\mathcal V_6}}  \,.
\ee

\noindent From this expression it is clear that  the range of $\chi$ depends on the geometrical parameters, encoded in  ${\mathfrak f}_0$ and $g^{\chi\chi}$ (and the compactification scale).
This is the effect of the warping in the WL inflationary scenario. In fact, in the flat slow roll case discussed in \cite{acq}, these two factors are not present.

\smallskip

To see explicitly how the fluxes and the warping enter into the normalisation of the WL field in (\ref{dimlesschi}) we need to evaluate
${\mathfrak f}_0$ in terms of generic geometric parameters.
As a guiding example we consider the Klebanov-Strassler \cite{KS} background. This geometry has generic features, which are also shared by more general warped backgrounds, as has been studied for example in \cite{hek,carlos}.

Let us start by looking  at the explicit form of  each  component of
\be\label{f0a}
f_0 = h_0^{-1} \mu_p\,g_s^{(p-3)/4} \int{d^{p-3}\xi\,
\sqrt{det(E_{0\,mn})}} \,.
\ee
As we have already mentioned, $\mu_p= (2\pi)^{-p}M_s^{p+1}$. Further, in  the KS solution (and in more general flux compactifications), the warp factor at a fixed position in the KS coordinates $\tau$ (see \cite{KS} for details) has the form
\be
h_0 \sim (g_s M \alpha')^2 \varepsilon^{-8/3} \cdot I_0 \,,
\ee
where $\varepsilon$ is related to the deformation of the conifold and has units
$\varepsilon^{2/3} = [Length]$, and $I_0$ is the integral of the KS solution evaluated at a fixed position, say $\tau_0$ (in the notation of \cite{KS}). The integral $I$ in the KS solution is a decreasing function of $\tau$ and has a maximum at the origin where it has a value  $I(0)\sim 0.71805$ \cite{KS}, thus $I_0\lesssim 0.71805$.
Remember also that $M$  are the units of  3-form flux and in the KS solution, this number is the same for both RR and NSNS 3-form fluxes.
Now remembering  that $E_{0\,mn} = h_0^{1/2} g_{mn} + g_s^{-1/2} (B_{mn} + 2\pi\alpha' F_{mn})$, we can
take $\alpha' F_{mn}$ of the same order of magnitude as $B_{mn}$.  Further, in the KS solution  $B_{mn}\sim g_sM\alpha' \varepsilon^{-4/3}\cdot (k_0 + \tilde k_0)$, where $k_0$ and $\tilde k_0$ are two functions of the coordinate $\tau$  \cite{KS}, which vanish at the tip of the cone,  $\tau =0$, and tend to $\tau/2$ as $\tau\to \infty$. Then we can write:
\be
E_{0\,mn} \sim \left(g_s^{1/2} M \alpha' \varepsilon^{-4/3}\right) [g_{mn} g_s^{1/2} +
             (k_0 + \tilde k_0 + f_2)] \,.
\ee
The determinant in (\ref{f0a}) will have the form
\be\label{determinantE}
\sqrt{\det (E_{0\,mn})} \sim  \left(g_s^{1/2} M \alpha' \varepsilon^{-4/3}\right)\,{\mathcal B}_0^{(p-3)/2}\,,
\ee
where we have defined $ {\mathcal B}_0 = [g_{mn} g_s^{1/2} + (k_0 + \tilde k_0 + f_2)] $. This can be taken to be of order $(k_0 + \tilde k_0)$, once the magnetic field of the same order.
Combining this with the definition of $h_0$ and $\mu_p$ we find
\be\label{f0string}
f_0 \sim \frac{(g_s M \alpha' \varepsilon^{-4/3})^{(p-7)/2}\,V_{p-3}}{(2\pi)^p \ell_s^{p+1}} \,\frac{{\mathcal B}_0^{(p-3)/2}}{I_0}\,,
\ee
where $V_{p-3} = \int{d^{p-3}\xi}$ is the volume of the internal $(p-3)$-cycle (this can include a wrapping number). Notice also that the value of the last term ${\mathcal B}_0^{(p-3)/2}/I_0$ can be very large or very small, depending on the actual  position at which it is evaluated.

Defining the dimensionless cycle volume
${\mathcal V}_{p-3} \equiv V_{p-3}/\ell_s^{p-3}$ and putting it back into the expression for $f_0$, we end up with
\be\label{f0size}
{\mathfrak f}_0 \sim  \frac{(g_s M)^{\frac{(p-7)}{2}} {\mathcal V}_{p-3} }{(2\pi)^p}
\left( \frac{\ell_s}{\varepsilon^{2/3}}\right)^{p-7} \,\frac{{\mathcal B}_0^{(p-3)/2}}{I_0}\,.
\ee

\noindent  Thus the geometrical dependence of the field in
(\ref{dimlesschi}) encoded in ${\mathfrak f}_0$, is given by the amount of flux (this coincides with that found in \cite{wrappedbranes}) as well as on the volume of the cycle wrapped by the brane. Besides, it depends on the quantity ${\mathcal B}_0^{(p-3)/2}/I_0$, whose value depends in turn, on the fixed location of the D-brane in the throat.

\smallskip

It is interesting to compare expression (\ref{dimlesschi}) with that obtained in the single field position DBI model. In that case, the range of the inflaton goes like
$(\Delta \varphi/M_{Pl})^2  \lesssim {\mathcal S} M^{(p-7)/2} $, where ${\mathcal S}$ is the normalisation factor, which for a D$3$-brane is just ${\mathcal S}=4$ \cite{bmca}, but it depends on the wrapping number and geometrical construction for $p=5,\,7$ \cite{wrappedbranes}.
Clearly for a position field $\varphi$ in the D$3$-brane case, the range of $\varphi$, and thus the length of the throat,  was constrained to be very small in order to be in agreement with
the approximations. On the other hand, the D$5$ and D$7$-branes gave a slightly better result \cite{wrappedbranes}. However in those cases, the backreaction caused by the high speed motion was under less control. Also,  cosmological constraints gave inconsistent results for all three types of standard position DBI inflation in \cite{LH,AL}.

\subsection{Cosmological Implications }

In this section we derive cosmological consistency constraints for warped Wilson line DBI inflation. As we shall see, these constraints are independent of the potential for the inflaton field, which we leave unspecified. They are also independent of the specific form of the background geometry, that is, the explicit form of ${\mathfrak f}_0$.
This is an important feature as it allows us to bound the theoretical parameter region where we can look for  potentially successful warped Wilson line inflation, in agreement with observations.  We will see in the last part of this section that there exist a small window of the parameter space where all constraints are satisfied.

\subsubsection{Cosmological Perturbations}

We start by rewriting the equations of motion in the single WL field in the Hamilton-Jacobi  formalism \cite{HJ}.  Differentiating (\ref{Fried}) with respect to time and using (\ref{fluid}) and (\ref{T00}-\ref{Tii}),
we find:
\be\label{chidot_of_gamma}
H^\prime=-\frac{\gamma}{2M_{Pl}^2}\, \dot\chi \, ,
\ee
and solving for $\dot\chi$:
\be\label{chidot}
\dot\chi = - \frac{2H^\prime}{\sqrt{M_{Pl}^{-4}+4f_0^{-1}
H^{\prime 2}}} \,,
\ee
\smallskip

\noindent where a $\prime$ means derivative with respect to the WL field.  We have explicitly used the dependence of $\gamma$ on $\dot\chi$ (eq.~(\ref{gamma})) and have kept the $(-)$ sign solution, in agreement with equation (\ref{chidot_of_gamma}).
Piecewise in regions where $\dot\chi$ is monotonic, we can use equation
(\ref{chidot}) to eliminate $\dot\chi$, and treat the scalar field
$\chi$ as the evolution variable to solve for $H(\chi)$.
Using (\ref{chidot}) we can also find an expression for $\gamma$ as:
\be\label{gammaHJ}
\gamma= \sqrt{1+ 4M_{Pl}^4\,f_0^{-1}\,H'^2}\,.
\ee
Plugging back (\ref{chidot_of_gamma}, \ref{chidot}) and (\ref{gammaHJ}) into the Friedmann equation we get
\be
V(\chi)= \frac{1}{3M_{Pl}^2} \,H^2 + qF_0 - V_{\varphi_0} - f_0  \sqrt{1+ 4M_{Pl}^4\,f_0^{-1}\,H'^2}\,.
\ee

Let us now have a look at the cosmological perturbation spectrum. For the present model, we can simply use the results of inflationary models with general kinetic terms \cite{GM} specialised to position  DBI inflation. In this case, the  inflationary parameters are given as follows \cite{GM,fnl}:
\bea
&& \epsilon \equiv  -\frac{\dot H}{H^2} = \frac{2M_{Pl}^2}{\gamma} \left(\frac{H'}{H}\right)^2\,,
					\label{epsilon}\\
&& \eta \equiv \frac{2M_{Pl}^2}{\gamma} \frac{H''}{H} \label{eta}\,,\\
&& s \equiv \frac{\dot c_s}{c_s H} = \frac{2M_{Pl}^2}{\gamma} \frac{H'}{H}\frac{\gamma'}{\gamma}\,,
				\label{s}
\eea
where in the second equalities we have used the relations  (\ref{chidot_of_gamma}, \ref{chidot}). Remember also that
\be\label{cs}
c_s = \gamma^{-1} =\sqrt{1-f_0^{-1}\dot\chi^2}\, .
\ee
Assuming slow roll conditions on these parameters during observable inflation, one can obtain the amplitudes and spectral indices for the scalar and tensor perturbations. They are given as follows \cite{GM}
\bea\label{spectra}
P_S^2 &=& \frac{1}{8\pi^2M_{Pl}^2}\,\frac{H^2}{c_s\,\epsilon} = \frac{H^4}{4\pi^2 \dot \chi^2}\,, \nn \\
P_T^2 &=& \frac{2}{\pi^2} \frac{H^2}{M_{Pl}^2}\nn\,, \\
1-n_s &=& 4\epsilon - 2\eta + 2s \nn\,, \\
n_t &=& -2\epsilon \nn \,, \\
r &\equiv& P_T^2/P_S^2 = -8c_s\,n_t = 16\,c_s\,\epsilon \,,
\eea
where the right hand side quantities are to be evaluated at the time of sound horizon crossing $kc_s=aH$ (where $k$ is the comoving wave number).
The observed normalisation of the CMB anisotropy power spectrum implies that
$P_S^2 = 2.5 \times 10^{-9}$. On the other hand, the recent WMAP5 data \cite{wmap5} indicate that
$n_s = 0.963^{+0.014}_{-0.016}$ if it is assumed that $r$ is negligible. If instead one assumes $r\ne 0$, the bound is given by (WMAP5 combined with BAO and SN data) $n_s = 0.969\pm0.012$. The upper limit on the scalar to tensor ratio from combined WMAP5+BAO+SN data is $r<0.25$ \cite{wmap5,AL}.

An important consequence of DBI inflation, is that the speed of sound is much smaller than the speed on light (\ref{cs}),  because the $\gamma$ factor has to reach very large values in order for the potential energy to dominate, producing  accelerated expansion. This implies a departure from purely Gaussian statistics which can be very large \cite{st}. These deviations are usually quantified in terms of the nonlinearity parameter, $f_{NL}$ \cite{st}. In the equilateral triangle limit, the leading order contribution to this parameter is given by \cite{fnl}
\be\label{fnl}
f_{NL}  = -\frac{1}{3}\left( \frac{1}{c_s^2} -1\right)\,.
\ee
This imposes a strong constraint on the speed of sound, or in other words, on the $\gamma$ factor. The maximum allowed value for $f_{NL}$, given by the most recent WMAP5 results, is $-151<f_{NL}<253$. Therefore, eq.~(\ref{fnl}) implies
\be\label{gammalimit}
 \gamma \lesssim 21\,.
 \ee
This is a restrictive bound for the ordinary DBI as well as for WLDBI, where in order to achieve enough acceleration, $\gamma$ needs to be very large.

\subsubsection{Lyth bound: upper bound on $r$}

Another important expression which relates the amount of tensor perturbations to the range of the inflaton field is provided by the Lyth bound  \cite{Lyth}, which we use here in the form proposed by \cite{LH}
\be\label{lyth}
 \left( \frac{\Delta\chi}{M_{Pl}}\right)^2 \simeq r \,\frac{(\Delta{\mathcal N})^2}{8}\, .
\ee
Here, $\Delta{\mathcal N}\lesssim 4$ corresponds to the number of e-folds while observable scales are generated \cite{LH}.
Combining  the field range bound (\ref{dimlesschi}) with the Lyth bound (\ref{lyth}), we find an upper bound on the gravitational wave production parameter in the present model. This gives
\be\label{upper}
r <  \frac{(2\pi)^{11}}{(\Delta{\mathcal N})^2}\,g^{\chi\chi} g_s\,
                 \frac{{\mathfrak f}_0}{l^2\,{\mathcal V_6}}  \,,
\ee
and it is given in terms of the geometrical quantity ${\mathfrak f}_0$ and the compactification scales $l$, ${\mathcal V_6}$.

\subsubsection{Lidsey-Huston bounds: lower  bounds on $r$}

We now derive two consistent lower bounds for any Wilson line DBI single field scenario, following Lydsey-Huston \cite{LH} (see also \cite{AL}). These are independent of the form of the potential, which can be left unspecified.
The lower bounds thus obtained can then be confronted with the upper bound coming from the Lyth relation (\ref{upper}) and with current observations $r<0.25$.
It is interesting to point out that the same analysis performed in the standard position DBI inflationary scenario gave rise to two inconsistent  bounds for $r$ for all types of branes, questioning the success of the single-field position DBI model \cite{LH,AL} (see however \cite{hltw}).
Let us see how these bounds arise in our scenario.

Using (\ref{cs}, \ref{chidot}) and the definitions of $\eta$ and $s$ (\ref{eta}, \ref{s}), we can rewrite the spectral index as follows \cite{LH}:
\be
 1-n_s = 4\epsilon + \frac{2s}{1-\gamma^2} \,.
\ee
Further, making use of  (\ref{fnl}) and $r$ in (\ref{spectra}), this can be re expressed as\footnote{This expression coincides with that obtained in \cite{LH} when their $\dot T=0$, as it should.}
\be
1-n_s = \frac{r}{4}\sqrt{1-3f_{NL}} + \frac{2s}{3f_{NL}}\,.
\ee
Assuming, as it happens in DBI, that $f_{NL}$ is large and negative, and that $s$ is very small, in agreement
with the slow roll conditions used to derive these results, we get that
\be
1-n_s < \frac{r}{4}\sqrt{1-3f_{NL}}\,.
\ee
From this relation, we obtain a lower bound on $r$ by using the maximum allowed value for $f_{NL}$, given by the most recent WMAP5 results, $-151<f_{NL}<253$ and those for $n_s$. This bound becomes
\be \label{lower}
 r \gtrsim 0.008 \,,
\ee
for $1-n_s= 0.043$.
As pointed out in \cite{AL} this is a strong restriction which sets a challenge to the models of standard position DBI inflation using also wrapped branes \cite{wrappedbranes}.

\smallskip

Let us now extract a second lower bound on $r$ for the WL DBI scenario and compare with the result in standard DBI.
Consider the definition of  $\epsilon$ in (\ref{epsilon}) in terms of $H'$. Using (\ref{chidot}) in this expression, we find
$$ H^2= \frac{\gamma\,\dot\chi^2}{2M_{Pl}^2\,\epsilon}\,.$$
Plugging  this and the relation for $r$ in (\ref{spectra}) into the expression for $P_S^2$ in eq.~(\ref{spectra}), we obtain
 $$ P^2_S = \frac{16}{\pi^2\,M_{Pl}^4}\frac{\dot\chi^2}{r^2}\,.$$
Noticing further that $\dot\chi^2 = (1-c_s^2)\,f_0$ and $c_s^2 = 1/(1-3f_{NL})$, we find an expression for $f_0$ in terms of cosmological parameters as follows \cite{LH}:

\be \label{constraintA}
\frac{f_0}{M_{Pl}^4} = \frac{P_S^2\,\pi^2 \,r^2}{16}\left(1- \frac{1}{3f_{NL}} \right) \,.
\ee

Thus, we can now use the Planck mass (\ref{planckmass}) in the expression above and combine it with (\ref{dimlesschi}) and (\ref{lyth}) to find a second lower bound, consistent with (\ref{lower})

\be\label{constriantB}
 r\gtrsim \frac{32\pi(\Delta{\mathcal N})^2}{P_S^2}\frac{l^2\,g_s^3}{g^{\chi\chi}{\mathcal V_6}}\,.
\ee
Note that this second bound is independent of $f_0$, but depends as well on the compactification scale.
Comparing with position DBI inflation,  the same line of reasoning above, gave rise to an upper bound $r < 10^{-7}$ \cite{LH}, thus in clear tension with the previous bound (\ref{lower}).
However, in the present case, as we have seen, the inflaton WL has a different origin and thus we obtained a second  lower bound, which is therefore  in accordance with (\ref{lower}).

\subsubsection{\small{Consistency of Bounds}}

In summary, in order for a warped DBI Wilson line inflationary scenario to be consistent with observational constraints, we have to look for solutions within the range of theoretical parameters, such that relations (\ref{upper}) and (\ref{constriantB}) are  in agreement with the observational current bound, i.~e.
\bea\label{allreqs}
&& r <  \frac{(2\pi)^{11}}{(\Delta{\mathcal N})^2}\,g^{\chi\chi} g_s\,
                 \frac{{\mathfrak f}_0}{l^2\,{\mathcal V_6}}  \,, \nonumber\\
&& r\gtrsim \frac{32\pi(\Delta{\mathcal N})^2}{P_S^2}\frac{l^2\,g_s^3}{g^{\chi\chi}{\mathcal V_6}} \,,\nonumber \\
&& r<0.25 \,.
\eea
It is not hard to see that the lower and upper bounds are not easy to achieve due to the correlations between the geometrical parameters,  requiring some amount of fine tuning. However,  there exists  a region of the parameters in agreement with all theoretical approximations, where the bounds above can be satisfied. Let us bound this region using our previous analysis of all constraints.

First of all, observe that in order to get large values of $\gamma$, so that potential energy domination is ensured, we need ${\mathfrak f}_0<1$. On the other hand, from the constraints above (\ref{allreqs}), we would like to keep ${\mathfrak f}_0$ as large as possible.
From the expression for ${\mathfrak f}_0$ in eq.~(\ref{f0size}), we see that for a D$7$-brane, it depends on the volume of the  internal four-cycle ${\mathcal V}_{4}$, and the value of the quantity ${\mathcal B}_0^{2}/I_0$ at a given position $\tau_0$. Instead, for a D$5$-brane, ${\mathfrak f}_0$ depends on the inverse of the flux, which gives a small contribution (as in \cite{wrappedbranes}) when $M$ is large (as required by the supergravity approximation). This can  be compensated with a large value of ${\mathcal B}_0/I_0$ and the volume of the two-cycle ${\mathcal V}_{2}$, in order to increase the value of ${\mathfrak f}_0$.
Next, the dependence on the six dimensional volume suggest that it cannot  take very large values. This is in accordance to the fact that we are not looking at a large volume approximation, where the effect of the warping becomes negligible. Now, in order for stringy effects to be negligible, we require all quantities to be larger than the string scale, that is $l>1$ and ${\mathcal V}_6^{1/6}>1$. Moreover, in general we have ${\mathcal V}_6>l^6$. Finally, for the string  perturbation expansion to be trustable, we require $g_s<1$ and for the supergravity approximation to hold,  and the backreaction effects to be small (see eq.~(\ref{backreaction1})), the amount of flux units has to be large $g_sM\gg1$.
Now, before taking specific values for the parameters, we need to remember also that, in a warped compactification, the six dimensional volume has a contribution coming  from the throat. In particular, we have ${\mathcal V}_6 \gtrsim {\mathcal V}_6^{th} $. In a KS type geometry, ${\mathcal V}_6^{th} \propto (g_sM)^2$. Thus, we need to check that the ${\mathcal V}_6$ and $ (g_sM)^2$ are taken consistently.
Once the compactification volume is fixed, the value of $\gamma$ has to satisfy (\ref{gammalimit}).
Finally, the value of $g^{\chi\chi}$ depends on the actual throat geometry.

 Given  all these considerations, for a D$7$-brane for example, the following set of values for the parameters, in string units, satisfies all required conditions:
 $g_s =10^{-2}, \,\, l= 4, \,\, {\mathcal V}_6 \sim 8\times10^5, \,\,{\mathfrak f}_0\sim 10^{-1}, \,\,
 {\mathcal B}_0^{2}/I_0\sim 60,   \,\,  {\mathcal V}_{4}^{1/4} \sim 5,\,\,
 g^{\chi\chi} \sim 5,  \,\,    \Delta{\mathcal N} =1,$   and produces a tensor spectrum in the range:

\be
0.16 < r <0.24 \, .
\ee

\bigskip

Therefore, warped DBI Wilson line inflation can potentially give rise to observable gravitational waves, as well as large detectable non-Gaussian perturbations.
Notice that the values of the parameters above depend on the exact position where the probe brane is sitting, $\tau_0$. In order to check that this can be achieved one needs  an explicit geometrical construction where the model can be embedded.
Still, it is remarkable that we can obtain two distinctive features from a string-motivated inflationary scenario.

\section{Conclusions\label{sec_concl}}

In this paper we have proposed that a Wilson line, as opposed to position degree of freedom,  can be responsible for driving  inflation in a DBI  fashion.
We derived in generality the DBI+WZ actions describing Wilson lines as well as D$p$-brane position fields. As we saw, once these two different types of fields are present, mixed kinetic terms arise. These can have important  consequences for multifield inflation led by Wilson line and position fields.
We coupled this general action to gravity, in order to study cosmology.
To single out the cosmological properties of Wilson line fields with non standard kinetic terms, we assumed that open string moduli associated to brane positions can be stabilised (e.g.~in the presence of 3-form flux using the techniques developed in \cite{GomisMarchMat}).

For concreteness, and in order to compare our DBI Wilson line scenario with the more studied position DBI models, we restricted our study to a single WL field. We determined the normalisation of the canonical WL fields and saw that, as expected, it picks up geometric and  flux dependent factors, in contrast to the flat slow roll version of Wilson line inflation \cite{acq}.

We determined consistency relations, both from theoretical as well as cosmological points of view.
From  the theoretical side, we pointed out the necessary restrictions on the parameters in order for the required approximations to be  valid. First, for the brane gravitational backreaction to be negligible, the probe brane limit requires large background flux values. The supergravity approximation also requires large values of flux, as these dictate the curvature of the ambient space, which needs to be small.
String perturbation theory implies that the string coupling be small, and the requirement that stringy effects
remain negligible implies that compactification scales in string units have to be larger than one.
A less trivial constraint comes from the backreaction due to the large values of the WL ``velocities". We estimated this effect in a simple way using T-duality, then giving the same type of bound as in the position DBI scenario \cite{egmtz}.
This gives a good estimate of this effect, but as we discussed, a more accurate calculation may be needed in a more realistic compactification with fluxes.

On the cosmological side, we derived bounds on the tensor to scalar  ratio, $r$,  independent of the form of the inflaton potential. Interestingly enough, given the different origin on the WL field as compared to positions, we found that DBI Wilson line inflation gives consistent bounds on $r$. Moreover, agreement of the Lyth bound combined with the Lidsey-Huston bound, allows a large value for $r$ and thus an observable gravitational wave spectrum. We showed  that there exist a range of values for the parameters, where the model can be realised. Taking into account all requirements, we presented an explicit set of values in the case of a D$7$-brane which can then produce large nonlinearities $f_{NL}$ as well as large gravitational waves $r$.

Recently, there has been some effort in trying to construct inflationary models within string theory, which can produce observable gravitational waves \cite{hltw,monodromy,CBQ}. It is thus remarkable that DBI Wilson line inflation provides yet another example where observable gravitational waves can be produced.
Even more interesting is the fact that warped WL DBI inflation produces  large non-Gaussianities as well. Therefore, if both gravitational waves and non-Gaussianities are observed, the models in \cite{monodromy,CBQ} will be excluded whereas WL DBI inflation could still survive.

Before ending, we point out that, just like  all models of brane inflation to date,
our scenario has some important challenges to overcome.
Perhaps the most obvious one concerns the construction of a realistic theoretical set up, where DBI WL inflation can be embedded, as we have already mentioned.
Given the two distinctive features of the present scenario, which we singled out in this paper, investigation on the theoretical side deserves further  attention. We plan to come back to this in the future.

Another challenge of the present scenario may be related to the observation in \cite{ms} (see also \cite{chen}) related to position DBI inflation. It  basically requires that the  backreaction caused by the effects which produce the potential for the position field, be small such that the geometry probed by the brane remains (almost) unchanged.  It is easy to see that the line of reasoning in \cite{ms} (and \cite{chen}) does not apply to the Wilson lines since they are not related to the geometry itself, or in other words, to the warp factor. At first sight, therefore, it is not obvious that the same problem would arise in the Wilson line DBI scenario, once position fields have been successfully stabilised.
 However, it remains to be seen whether  position fields can be stabilised without affecting the Wilson line moduli.  Clearly to address this and other issues, we  require a more concrete realisation of the present set up. One could imagine already a toy model in five dimensional brane world model \'a la RS \cite{RS}, along the lines of \cite{oda}, where some of these questions can be answered \cite{progress}.

Further to this, there are several avenues that can be explored in warped Wilson Line DBI inflation. Perhaps the most natural one regards multifield inflation.  In fact, in a more realistic stringy inflationary scenario, we expect more fields to play a role in the cosmological evolution of the universe.  For example, brane positions as well as Wilson lines could both drive inflation, and as we have seen mixed kinetic terms can arise. Multifield DBI inflation has been studied recently in \cite{egmtz,shiu,langlois}. It would be interesting to use these results to combine position and WL DBI scenarios. We leave these and other open issues  for future investigation.

\section*{Acknowledgements}
We would like to thank R.~Gregory, Y-H. He, C.~N\'u\~nez, F.~Quevedo,  A.~Uranga and especially G.~Tasinato  for useful discussions.  AA is supported by the EC Marie Curie Research Training Network ENRAGE. This work is also supported in part by MEC, research grant FPA2007-66665.  IZ was supported in part by an STFC Postdoctoral Fellowship and in part by  the European
Union 6th framework program MRTN-CT-2004-503069
``Quest for unification", MRTN-CT-2004-005104 ``ForcesUniverse", MRTN-CT-2006-035863 ``UniverseNet" and SFB-Transregio
33 ``The Dark Universe"
by Deutsche Forschungsgemeinschaft (DFG).
AA thanks IPPP for hospitality while this work was initiated.

\end{document}